\documentclass[reprint,
superscriptaddress,
frontmatterverbose, 
amsmath,amssymb,
aps,
prb,
floatfix,
showkeys
]{revtex4-2}

\usepackage{float}
\usepackage{graphicx}
\usepackage{dcolumn}
\usepackage{bm}
\usepackage[left=2cm,right=2cm,top=2cm,bottom=2cm]{geometry}
\usepackage{mathtools}
\usepackage{amsmath}
\usepackage{lipsum}
\usepackage{xcolor}
\usepackage{soul}
\usepackage{multirow}
\mathchardef\hyphen="2D

\begin{document}

\preprint{APS/123-QED}

\title{Thermal rectification and phonon properties in partially perforated graphene.}

\author{Markos Poulos}
 \email[Corresponding Author: ]{markos.poulos@insa-lyon.fr}
 \affiliation{INSA Lyon, CNRS, CETHIL, UMR5008, 69621 Villeurbanne, France}

\author{Konstantinos Termentzidis}
\affiliation{CNRS, INSA Lyon, CETHIL, UMR5008, 69621 Villeurbanne, France}

\begin{abstract}
In this work, a thermal rectification ratio $\eta$ of 18.5\% was observed in partially perforated graphene with the use of Molecular Dynamics (MD) simulations. In all cases studied here, heat preferentially flows from the porous to the pristine region and both $\kappa$ and $\eta$ increase upon increasing the length of the pristine region and upon decreasing the size of the pores. To interpret the results, the macroscopic “R-Series Model” is applied, attributing rectification to the different temperature-dependence of $\kappa$ of perforated and pristine graphene. According to the model, $\eta$ is maximized when the two regions composing the structure have matching thermal resistances and mismatching temperature-dependence of $\kappa$. The model agrees qualitatively with the MD results, indicating that the latter is the principal rectification mechanism, but it can significantly underestimate $\eta$. Phonon analysis further reveals the appearance of new `\textit{defect}' modes localized around and between pores, resulting in the emergence of a new prominent peak in the phonon Density of States at 520 cm$^{-1}$. The study considers key geometric factors such as the length of the pristine region, and the pore size, shape, alignment, and orientation. Pore shape and alignment exert minimal influence on $\eta$, although alignment greatly influences $\kappa$. Eventually, arranged pores are deemed more efficient than randomly distributed defects for increasing rectification.
\end{abstract}

\maketitle

\section{\label{sec:Intro}Introduction}

Efficient thermal management is a critical challenge in modern nanotechnology, particularly for applications in electronics, energy conversion, and cooling~\cite{Ogrenci}. The concept of thermal rectification, wherein heat conduction is directionally dependent, holds promise for the development of thermal devices analogous to electrical diodes. Such devices can regulate heat flow, mitigate overheating, and optimize energy usage in nanoscale systems~\cite{WehmeyerReview, LiuReview, WongReview}. As this topic attracts a high technological interest, there have been numerous studies on building a functioning thermal rectifier, both experimentally as well as theoretically and computationally. Although thermal rectification has been long demonstrated using bulk materials~\cite{Balcerek1978, Jezowski1978}, intense reaserch focus has been placed in nanoscale systems, using either 3D nanostructures~\cite{Lee2012,Han2021,Wei2019, Cartoixa2015}, more recently using using low-dimensional material latteral heterostructures, involving graphene and other low dimensional materials~\cite{Desmarchelier2021, Chen2020,Hu2009,Wang2014,Zhong2011,Zhang2022,Chen2019,Shrestha2020,Tavakoli2022,Yang2017,Zhao2015}.

Among the potential candidates for thermal rectifiers, graphene-based materials have emerged as a focal point due to their exceptional thermal and mechanical properties~\cite{Papageorgiou2017,Lee2008}. Graphene, a two-dimensional material composed of a single layer of carbon atoms arranged in a hexagonal lattice~\cite{Geim2007}, exhibits an extraordinarily high thermal conductivity, surpassing most other materials at room temperature~\cite{Nika2017}. The thermal properties of graphene have been the subject of numerous studies in the past decade, due to its potential applications in thermal management, thermoelectricity and thermal rectification~\cite{Hu2009,Wang2014,Zhong2011}. The thermal conductivity of pristine graphene has been found to be very high, with values reaching up to 5000 W/mK at room temperature~\cite{Balandin2008}, while it is also highly temperature- and length-dependent, with some studies suggesting that it reaches the diffusive plateau (independence on system length) at lengths of the order of $\sim$100 $\mu$m $-$ 1 mm~\cite{Barbarino2015,Fugallo2014}. The importance of including 4-phonon processes~\cite{Han2023} and the dominance of the Normal scattering processes~\cite{Fugallo2014} in obtaining accurate results for $\kappa$ of pristine graphene has also been demonstrated.

However, the thermal conductivity of graphene can be significantly reduced by introducing defects, such as vacancies, grain boundaries, or by creating porous structures. The latter have been shown to exhibit a wide range of thermal conductivities, depending on the porosity, pore shape, pore size, and pore distribution~\cite{Zhang2022}. However, uniquely pristine or uniquely porous graphene lacks an inherent thermal rectification effect due to symmetry. To achieve rectification, introducing structural asymmetry into the thermal transport pathway is thus essential. One promising strategy to induce this asymmetry is by combining pristine and porous graphene in a two-dimensional latteral heterostructure. These porous structures can have varying porosity, pore size, and pore distribution, and some previous studies, both experimental and computational, on these kinds of heterostructures have shown promising results ($\eta$= 5-70\%)~\cite{Yousefi2020,Arora2017,Chen2022,Nobakht2018}, and have been even experimentally tested ($\eta_{max}$= 26\%)~\cite{Wang2017}. As will be further elaborated subsequently, this structural asymmetry is a necessary but not sufficient condition for rectification.

Thermal rectification is a general phenomenon that arises from the asymmetric flow of phonons, which are the primary heat carriers in non-metals, across a system, when changing the direction of the thermal bias. Although there is so far no greater concensus in the litterature concerning the exact generic mechanism of thermal rectification, it has been proven that there are at least two necessary conditions: asymmetry along the heat flux direction and non-linearity of the thermal properties with respect to the temperature bias $\Delta$T~\cite{WehmeyerReview,Maznev2013,Baowen2004}

Various microscopic mechanisms have been proposed to explain the origins of thermal rectification and we shall hereby mention some of them. It shall be noted that although the following refer mostly to low-dimensional systems, the conclusions are general. In lateral heterostructures, the two lateral regions have in principle different phononic spectra, reflected in a mismatch in the phonon Density of States (DOS), while the interface perpendicular to the heat flux serves as a critical region for phonon scattering and transmission. The change of the DOS mismatch with reversing the temperature bias $\Delta$T is then held responsible for a differential scattering of phonons accross the interface, leading to rectification. This approach assumes two important approximations. First, that the modification of phonon transmission with flipping $\Delta$T, and thus the rectification, is only due to the modification of the phonon DOS with temperature~\cite{Chen2020, Chen2019,Baowen2004, Baowen2005}. Second, it neglects the appearance of new interfacial modes that do not exist in the consituting individual materials of the heterostructure. This is a crucial point, as it has been shown that the interaction of these new modes with the pristine modes can greatly affect conduction accross the interface~\cite{Henry2016,Henry2017}

These points are more adequately addressed by an interface-driven mechanism, where the rectification is due to the dependence of the interfacial phonon transmission on the interface temperature which can be caused by the presence of localized phonon modes, or by the change of the phonon transmission coefficients themselves with temperature. It has been shown that the interface can be described as an autonomous thermodynamic system, posessing a well-defined temperature T$_s$, and that the interfacial thermal resistance depends on T$_s$~\cite{Rurali2016}. In a heterostructure of materials with dissimilar thermal resistances, possessing a well-defined interface, T$_s$ at steady-state can be different depending on the direction of $\Delta$T, thus leading to rectification. Heterostructures of this kind have been shown to produce high rectification ratios $\eta$= 5-90\%~\cite{Zhang2022,Xu_TR2014,Rurali2018}.

It has also been suggested that a rectification can also arise due to asymmetric scattering of phonons in the ballistic/hydrodynamic regime, and rectification ratio as high as 300\% was demonstrated computationally in a bulk system with pyramidal nano-voids using a Landauer-B\"uttiker formalism~\cite{Miller2009}. Low temperatures and large $\Delta$T were used. A crucial point underlined in this study was the use of the displaced Bose-Einstein distribution function~\cite{Callaway1959} for the description of the heat flux out of equilibrium, as opposed to the ususal equilibium Bose-Einstein distribution used in the Landauer formulas for thermal conductivity. This point has also been known to affect the correct calculation of $\kappa$ and the Interfacial Thermal Resistance (ITR)~\cite{Merabia2012}.

However, rectification does not necessarily require microscopic asymmetry, since it can also be achieved by a macroscopic one. In one such model, which we shall henceforth call the R-series model, the driving force for the rectification is the different dependence of the thermal conductivity on temperature of the materials composing the heterostructure. The terminology was chosen, for lack of a proper name in the litterature, to reflect the analogy with the electrical equivalent, as the model assumes that the total system consists of two thermal resistances connected in series and $R_{total}=R_1+R_2$. The ITR often observed in latteral heterostructures can also be incorporated in the model~\cite{Rurali2018}. Ideally, one material should have a thermal conductivity that increases with temperature, while the other should have a thermal conductivity that decreases with temperature. This way, the heat flux will be higher when the material with the $\kappa$ increasing with $T$ is on the hot side, and lower when it is on the cold side~\cite{Dames2009}. The R-series model also neglects the effect of the interface, both in $R_{total}$, as well as how the appearance of the new interfacial modes affects the phonon dynamics of the pristine modes. It is thus much more adapted to macroscopic systems, where the effect of the interface can be neglected. Nonetheless, we shall apply this model to our microscopic systems and explain that a significant part of the rectification observed can be explained by this macroscopic mechanism. This model lies on an accurate description and an efficient engineering of $\kappa(T)$ for the constituent materials. In general, $\kappa$ can be given by the well-known Boltzmann-Peierls formula $\kappa = \sum_j c_{v,j} v_j^2 \tau_j  \label{eq:Phonon_Gas}$~\cite{Ziman}, where $c_{v,j}$ is the specific heat capacity, $v_j$ is the group velocity and $\tau_j$ is the relaxation time of the phonon mode $j$. The $T$-dependence of $\kappa$ is governed by the interplay of various microscopic mechanisms: increase in the phonon population with $T$ up to the Debye temperature $T_D$ ($\kappa\propto c_{v,j}\propto T^3$), Umklapp phonon-phonon scattering ($\kappa\propto\tau_j \propto T^{-n}$ ), boundary and defect scattering ($\kappa$ independent of $T$)~\cite{Ziman}. The dominant mechanism thus determines the total dependence $\kappa(T)$, which can often be approximated by a power law $\kappa=\kappa_0\left(\frac{T}{T_0}\right)^n$ in specific temperature regions. In the case of perforated graphene, the pores introduce additional boundary scattering which not only reduces the total $\kappa$, but also modifies its temperature dependence. In Section~\ref{sec:Method} C we briefly discuss the basic equations that serve to predict TR based on this power law.

The article is organized as follows. Section II details the methodology of the computational techniques used for the theoretical calculations, including a brief description of the various systems used, the Non-Equilibrium molecular Dynamics (NEMD) method for the direct calculation of $\kappa$ and R\%, as well as the macroscopic R-series model. Section III presents the results of the calculations of $\kappa$ and R\%, and discusses the comparison of the results with the R-series model, for a number of different configurations. Finally, Section IV highlights the most important findings and some concluding remarks.

\section{Methodology \label{sec:Method}}
\subsection{System Structures}
\begin{figure}[tbp]
    \begin{center}
        \includegraphics[width=1\columnwidth]{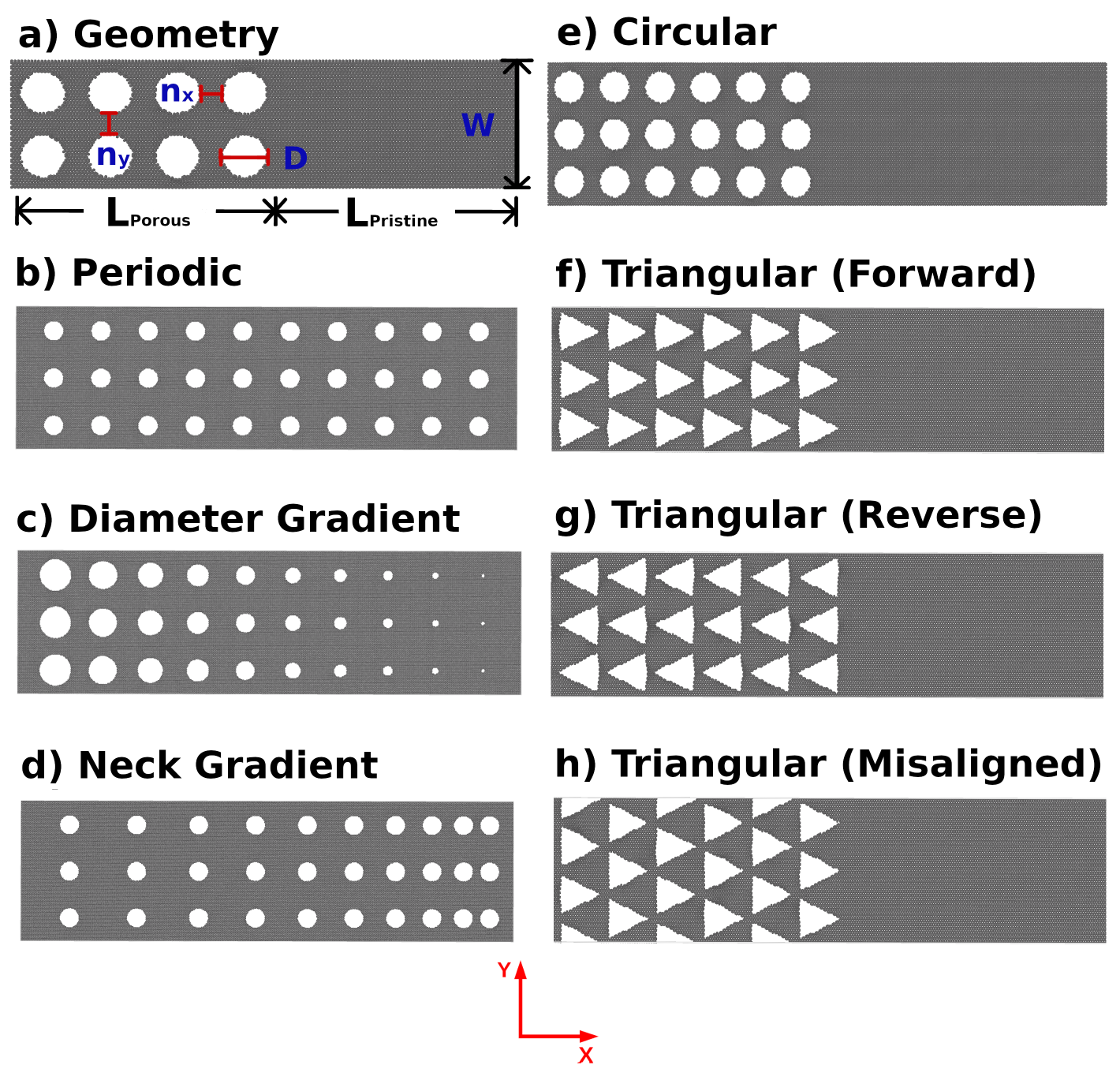}
        \caption{\label{fig:Config} (Right) Schematics of the perforated graphene configurations treated in this work. (a) Visual representation of the geometric parameters of the systems studied, like the pore diameters $D$ and necks ($n_x$, $n_y$), the lengths of the porous and pristine sides ($L_{porous}$ and $L_{Pristine}$), and the system width $W$.  Porous graphene with circular pores arranged (b) periodically, (c) with a gradient in diameters and (d) gradient of necks along the flux direction. (Left) Partially perforated graphene structures. (e) Circular pores. For the triangular pores, depending on the orientation of the triangle apex with respect to the pristine region, the pores are labeled (f)\textit{`Forward'} or (g)\textit{`Reverse'}. (h) Misaligned (forward) pores in subsequent rows.}
    \end{center}
\end{figure}

The systems studied were rectangular graphene monolayer sheets of constant width $w$ with pores of various sizes, shapes and arrangements. The two important geometric parameters are the pore diameters $D$ and the pore necks $n$ ($n_x$ and $n_y$ along the $x$- and $y$-axis respectively). The $n$ is defined as the minimum distance between the edges of two adjacent pores along each direction. For the triangular pores, which were made equilateral, $D$ was defined as the triangle height. The period of the pattern is then $\left(D+n\right)$. Another parameter is the porosity $p$, which can be defined as the ratio of the surface of all the pores to the total surface of the porous region, and is globally a function of $\left(\frac{D}{n}\right)$. The perforated graphene systems were combined with pristine graphene regions in order to create `lateral heterostructures', whose potential difference in the temperature dependence of $\kappa$ could lead to thermal rectification. Examples of all the studied configurations can be found schematically in Fig.~\ref{fig:Config}.

For the study of thermal rectification of partially perforated structures, we explored systems of constant width $w$= 50~nm, comprising a porous region of constant length $L_{porous}$= 50~nm, pore necks $n_x$=$n_y$=$n$ and various porosities $p$, as well as a pristine region of length $L_{pristine}$, varying from 0 to 450~nm. The pores geometries explored had a constant period of 5~nm and were of circular and triangular shape. The circular pores had $D$=1, 1.82 and 4~nm, necks of $n$=4, 3.18 and 1~nm, and porosities $p$= 15\%, 32\% and 50\% respectively. The triangular pores with $D$= 4~nm and $n$= 1~nm had $p$= 32\%. Note that due to their geometry, circular and triangular pores with $D$=4~nm and $n$=1~nm had different porosity (50\% and 32\% respectively). 

In order to examine a possible effect of the orientation of the triangles on the rectification ratio, we explored two different orientations of the apexes with respect to the pristine region. These orientations, with the apexes facing towards the pristine region, hereby referred to as \textit{`forward'}, and the apexes facing away, hereby referred to as \textit{`reverse'} are depicted in Figs.~\ref{fig:Config}f and~\ref{fig:Config}g. We also explored the effect of the alignment of the pores in subsequent rows, for the \textit{`forward'} orientation, with the pores being either aligned or misaligned (Fig.~\ref{fig:Config}h). We finally attempted to induce rectification through an asymmetrical pore arrangement, without the use of pristine regions (Fig.~\ref{fig:Config} (b,c,d)). However, no rectification was found above the uncertainty. Details and results for these configurations are provided in the Supplementary Material, as they were not the main focus of our study.

\subsection{Phonon Analysis with Lattice Dynamics}
Although graphene is a semimetal, it is known that phonons are the principal carriers of heat, with the electronic contribution to $\kappa$ being only $\sim$1\% of the total~\cite{Lindsay2010,Yigen2013}. The phonon modes of the systems studied in this work were calculated by means of Lattice Dynamics (LD). All vibrational modes were calculated by diagonalizing the full dynamical matrix of the system
\begin{eqnarray}
    D_{ij}^{\alpha \beta}=\frac{1}{\sqrt{m_i m_j}}\frac{\partial^2 U }{\partial u_{i}^{\alpha} \partial u_{j}^{ \beta}},  \label{eq:Dyn_Mat}
\end{eqnarray}
where the indices $i$,$j$ refer to atoms and $\alpha$ and $\beta$ are cartesian indices, $U$ refers to the total system potential energy and $u_i^{\alpha}$ is an infinitesimal displacement along the $\alpha$-axis of atom $i$. In order to simulate an infinitely large system, a computational cell containing $N$ atoms with Periodic Boundary Conditions (PBC) is considered.  Upon diagonalization of $D_{ij}^{\alpha \beta}$, one obtains 3$N$ modes with eigenfrequencies $\omega^2_k$ and their corresponding eigenvectors $\vec{e}_k$, $k$=1, 2 $\ldots$ 3$N$. The phonon Density of States (DOS) is then simply calculated by its definition as:
\begin{eqnarray}
D\left(\omega\right)=\sum_k \delta \left(\omega - \omega_k\right), \label{eq:DOS}
\end{eqnarray}
An important property of phonon modes is their degree of spatial localization, which can serve as a measure of the degree of participation of a phonon mode to thermal transport. This property is quantified by the phonon Participation Ratio (PR), which is defined for a mode $k$ by the degree of localization of its eigenvectors as~\cite{Bodapati2006}:
\begin{eqnarray}
    p_k^{-1}=N\sum_i \left[\vec{e}_{i,k}\cdot\vec{e}_{i,k}^{\ *}\right]^2, \label{eq:PR}
\end{eqnarray}
where $^*$ denotes complex conjugation, $i$=1, 2, 3 $\ldots$ $N$, is the atom index and $k$=1, 2, 3 $\ldots$ 3$N$ is a phonon mode index.

\subsection{NEMD Calculations}
The thermal properties of the porous/pristine graphene lateral heterostructures were calculated using the Non-Equilibrium Molecular Dynamics (NEMD) method. The orientation of our system was such that heat current flows along the zig-zag direction, which is aligned to the $x$-axis. We imposed PBC in the $y$ direction, and Fixed Boundary Conditions (FBC) in a few layers of atoms next to the edges, along the $x$-axis. For all calculations, all systems were first structurally and thermally equilibrated at an average temperature $T_0$=300 K. The LAMMPS~\cite{LAMMPS} software package was used for the simulations, with the reparametrized Tersoff potential by Lindsay and Broido~\cite{Broido2010} for the carbon-carbon interactions.

The thermal rectification ratio $\eta$ has been subject to many definitions, which give rise to different ranges of values~\cite{WongReview}. In this work, we use the following definition for the thermal rectification ratio $\eta$:
\begin{eqnarray}
    \eta = \frac{J_{+} - J_{-}}{J_{-}} \label{eq:Thermal_Rectification}
\end{eqnarray}
where $J_{+}$ ($J_{-}$) is the larger (lower) heat flux towards the easy (hard) direction, respectively. Using this definition, a perfect rectifier would have $\eta= \infty$, while absence of rectification is equivalent to $\eta=0~\%$. To obtain a single value of $\eta$ for each configuration, we repeat the NEMD calculation 5 times for each direction with different initial conditions, totalling 10 simulations of 10 ns each for every configuration. This way we were able to have an estimation of the uncertainties, which were typically of the order of 0.5\% of the calculated values. It should be noted that a single simulation per direction would not be enough to obtain a reliable value of $\eta$, as the differences in the heat fluxes can fluctuate across realizations with different initial conditions.

\subsection{The R-Series Model}\label{subsect:R_Series}
To obtain a theoretical prediction of the thermal rectification ratio $\eta$ of the systems studied with the R-series model, we first need to determine the temperature dependence $\kappa(T)$ of the thermal conductivities of the materials composing the heterostructures. The NEMD results were then fitted to the power law
\begin{eqnarray}
    \kappa(T) = \kappa_0 \left(\frac{T}{T_0}\right)^m  \label{eq:Power_Law}
\end{eqnarray}
where $\kappa_0$ is the thermal conductivity at a reference temperature $T_0$, here taken as $T_0$=300 K and $m$ an exponent that can be either positive or negative. After having obtained $\left(\kappa_0, m\right)$ for a given pair of materials 1 and 2 comprising the heterostructure, following Dames we can calculate the reduced heat current $q$, by solving the system of equations~\cite{Dames2009}:

\begin{subequations}\label{eq:R_Series_Q}
    \begin{align}
        q & = \left(\frac{1+\rho}{m_1+1}\right) \left(\theta_L^{m_1+1}-\theta_j^{m_1+1}\right) \\[1em]
        q & = \left(\frac{1+\rho^{-1}}{m_2+1}\right) \left(\theta_j^{m_2+1}-\theta_R^{m_2+1}\right) 
    \end{align}
\end{subequations}
where $\theta$=$\left(T/T_0\right)$ is the reduced temperature and the subscripts $L$, $R$ and $j$ refer to the temperatures of the left thermostat, the right thermostat and the junction respectively. $m_1$ and $m_2$ are the power law exponents of eq.~(\ref{eq:Power_Law}) for materials 1 and 2 respectively, and $\rho$=$R_{0,1}/R_{0,2}$ is the ratio of the thermal resistances $R_0$ of the two materials at $T$=$T_0$. Here, the two materials are assumed to have a common cross section $A$ and lengths $L_1$ and $L_2$. Then $R_0$=$L/\left(\kappa_0 A\right)$, and $\rho$ is given by
\begin{eqnarray}
    \rho = \left(\frac{L_1}{L_2}\right) \left(\frac{\kappa_{0,2}}{\kappa_{0,1}}\right) \label{eq:Rho}
\end{eqnarray}

Eqs.~(\ref{eq:R_Series_Q}) are a system of non-linear equations with known parameters $m_1$, $m_2$ and $\rho$ determined from the fitting of the NEMD $\kappa(T)$ with eq.~(\ref{eq:Power_Law}) and $\theta_L$ and $\theta_R$ from the imposed temperature bias $\Delta T$. The system has two unknowns, $q$ and $\theta_j$, and it can be solved numerically for the reduced heat current $q_{\pm}$ corresponding to $\pm \Delta T$. The rectification ratio $\eta$ is then calculated as in eq.~(\ref{eq:Thermal_Rectification}).

A noteworthy point is the importance of the estimation of uncertainties in the final results for $\eta$, which stem from the uncertainties in the fittings parameters $\left(\kappa_0, m\right)$ of eq.~(\ref{eq:Power_Law}), as well as the uncertainties in the NEMD calculations of $\kappa(T)$. The latter are taken into account by using a weighted non-linear least squares fitting algorithm for the fitting and reflected in the uncertainties of the fitting parameters $\left(\kappa_0, m\right)$. The propagated uncertainty to $q_{\pm}$ is estimated by solving the system of eqs.~(\ref{eq:R_Series_Q}) by a Monte-Carlo scheme and the final uncertainty of $\eta$ is then estimated by propagating the uncertainties in $q_{\pm}$ through eq.~(\ref{eq:Thermal_Rectification}).

\begin{figure}[!tbp]
    \begin{center}
        \includegraphics[width=0.8\columnwidth]{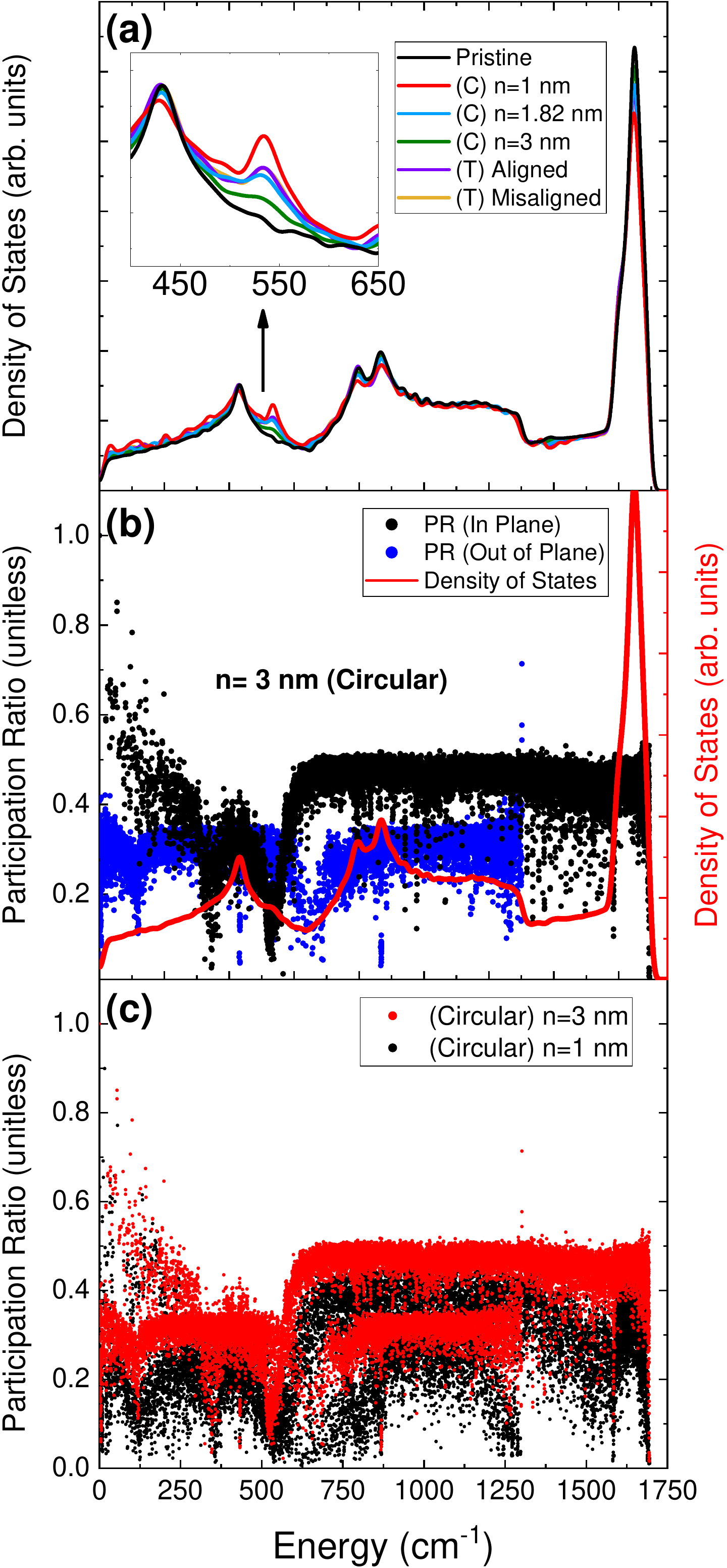}
        \caption{\label{fig:DOS_PR} The effect of perforation on the phonon properties of graphene. (a) The phonon Density of States (DOS) of pristine and porous graphene with circular pores (C) with necks $n$=1, 1.82 and 3~nm, and triangular pores (T) with aligned and misaligned pores. The inset is a magnification of the region around the new peak at 520 cm$^{-1}$. (b) The DOS and PR as a function of energy for the phonon modes of (C) $n$=3~nm of (a),  resolved for polarization: in-plane (black dots) and out-of-plane modes (blue dots). (c) PR as a function of energy for the phonon modes of system (C) $n$=3~nm (black dots) and (C) $n$=1~nm of (a) (red dots), respectively.}
    \end{center}
\end{figure}

\begin{figure*}[!tbp]
    \begin{center}
        \includegraphics[width=0.8\textwidth]{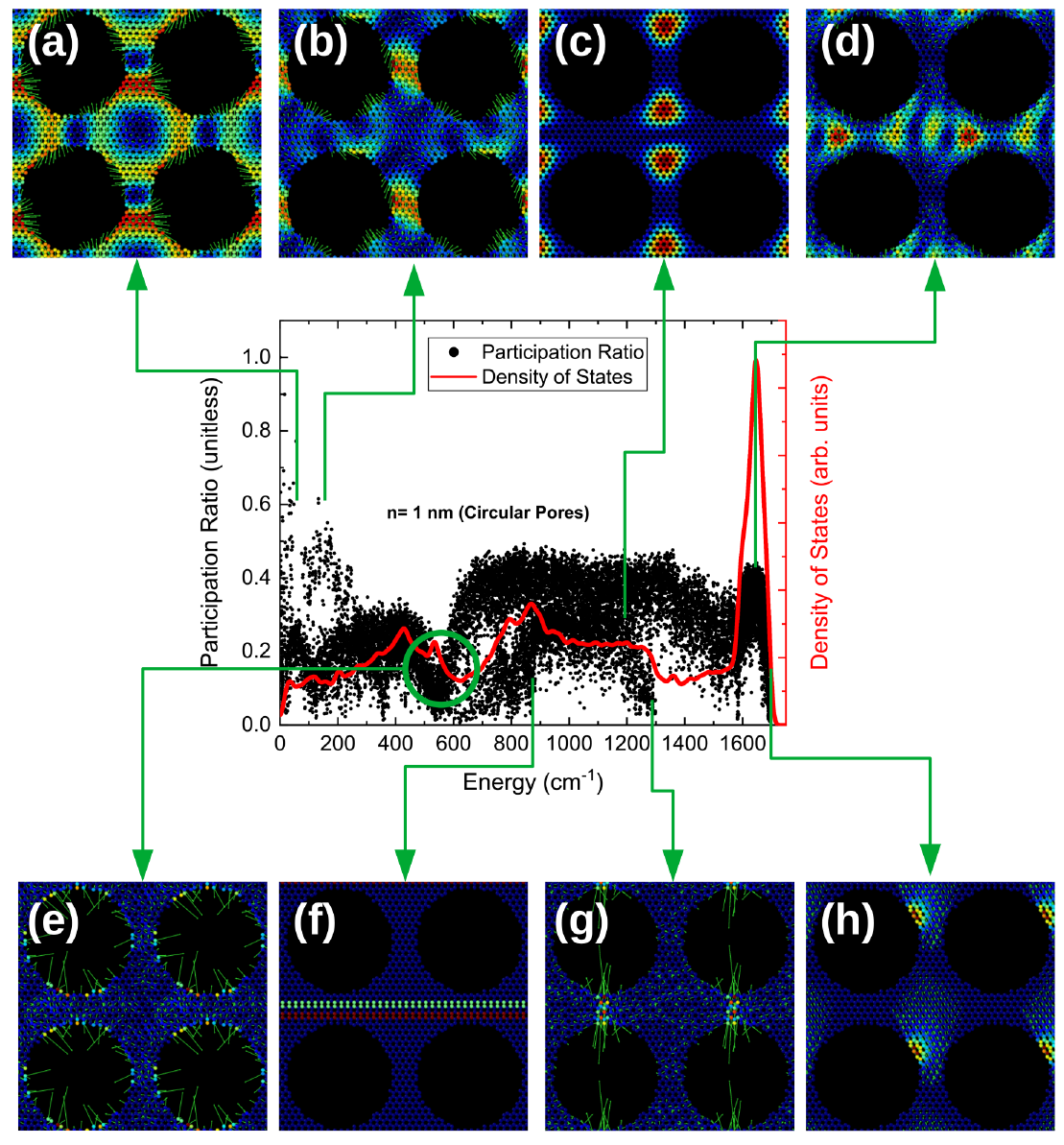}
        \caption{\label{fig:Modes} Visual representation of some characteristic phonon modes of perforated graphene with circular pores and $n$=1~nm, associated to their D($\omega$) and $p_k(\omega)$ values via arrows. (a-d) Extended modes, having a high $p_k$, and corresponding to pre-existing modes of graphene. (Middle) DOS and $p_k$ for the same system. The new peak at 520 cm$^{-1}$ is clearly visible. (e-h) Localized modes, concentrated around or between pores and corresponding to new `defect' modes. The modes associated with the 520 cm$^{-1}$ peak have a clear vibration pattern where under-coordinated pore edge atoms vibrate radially. In all cases the green arrows represent the mode eigenvectors $\vec{e}_k$ and the color coding is a measure of $|\vec{e}_k|$. In cases where the arrows are not visible, they correspond to out-of-plane modes, where the arrow axis is perpendicular to the page.}
    \end{center}
\end{figure*}

\section{\label{sec:Results}Results and Discussion}

\subsection{Phonon Properties of Porous Graphene}\label{subsect:Phonons}
We begin our discussion by analyzing the impact of perforation on the phonon properties of graphene. To this end, we performed lattice dynamics (LD) calculations on both pristine and perforated graphene for all the studied configurations. In all cases, the system dimensions were set to $L = 20$~nm and $W = 15$~nm, which provided enough spectral resolution in DOS and $p_k$ for the purposes of this work. These calculations allowed us to obtain the phonon Density of States $D(\omega)$ as well as the phonon Participation Ratio as a function of mode energy, $p_k(\omega)$, with the key results presented in Figure~\ref{fig:DOS_PR}. The introduction of pores leads to modifications in the phonon spectrum. The vibrational pattern of existing modes is modified, as can be seen Fig.~\ref{fig:Modes}a-d. More specifically, Figs.~\ref{fig:Modes}a-b correspond to high $p_k$, low energy, large $\lambda$ in-plane modes, while Figs.~\ref{fig:Modes}c-d correspond to high $p_k$, large $\lambda$ optical ZO and LO/TO modes respectively. More interestingly, new `\textit{defect}' modes appear, which are highly localized either around or between pores, as seen in Figs~\ref{fig:Modes}e-h. These modes can be identified by their very low $p_k(\omega)$ values and appear as vertical bands in the $p_k$($\omega$) plots at approximately 10, 120, 350, 550, and 1250 cm$^{-1}$ (Fig.~\ref{fig:DOS_PR}b). Some of these localized defect modes are illustrated in Figure~\ref{fig:Modes}e-h.

Most importantly, we observe a very large number of defect modes at around 520 cm$^{-1}$, which are responsible for the appearance of a new peak in the DOS at this energy, as seen in Fig.~\ref{fig:DOS_PR}a and Fig.~\ref{fig:Modes} (Middle). After visual inspection of $\vec{e_k}$ of these modes, we found that it corresponds to a radial vibration of under-coordinated atoms at the pore edges, as shown in Fig.~\ref{fig:Modes}e. This peak is a clear signature of the presence of pores in the system, and its intensity increases with the porosity of the system. The inset of Figure~\ref{fig:DOS_PR}a shows a magnification of the region around the peak, where the DOS of the perforated graphene with circular pores and $n$=1-3~nm are compared to the pristine system and a clear increase of the peak is observed as the pores become bigger ($n$ decreases). This result also indicates that the number of these modes is related to the percentage of pore edge atoms, and thus to the porosity of the system.

Additionally, our analysis reveals a clear separation between in-plane and out-of-plane vibrational modes, with a higher degree of localization observed for the out-of-plane modes, as evidenced by Figure~\ref{fig:DOS_PR}b, where most in-plane modes have an average $p_k\sim$0.5, while the out-of-plane ones have on average $p_k\sim$0.3, for the case of circular pores with $n$=3~nm. Moreover, an increase in porosity generally results in a global decrease in PR, indicating a higher degree of phonon localization (see Fig.~\ref{fig:DOS_PR}c), which is a more or less expected outcome. Interestingly though, the specific shape and alignment of the pores appear to be largely irrelevant in determining the Density of States (DOS); rather, the porosity itself is the dominant factor, as can be seen from Fig.~\ref{fig:DOS_PR}a, where all the DOS of all systems with triangular pores ($n$=1~nm, $p$=32\%) are almost identical to the DOS of the one with circular pores and $n$=1.82~nm (having the same porosity $p$=32\%). This observation also suggests a strong correlation between phonon properties and the ratio of edge-to-surface atoms, and by extension, the porosity of the system.

\subsection{Thermal rectification via partial perforation}\label{subsect:Heteros}

\begin{figure}[!tbp]
    \begin{center}
        \includegraphics[width=0.8\columnwidth]{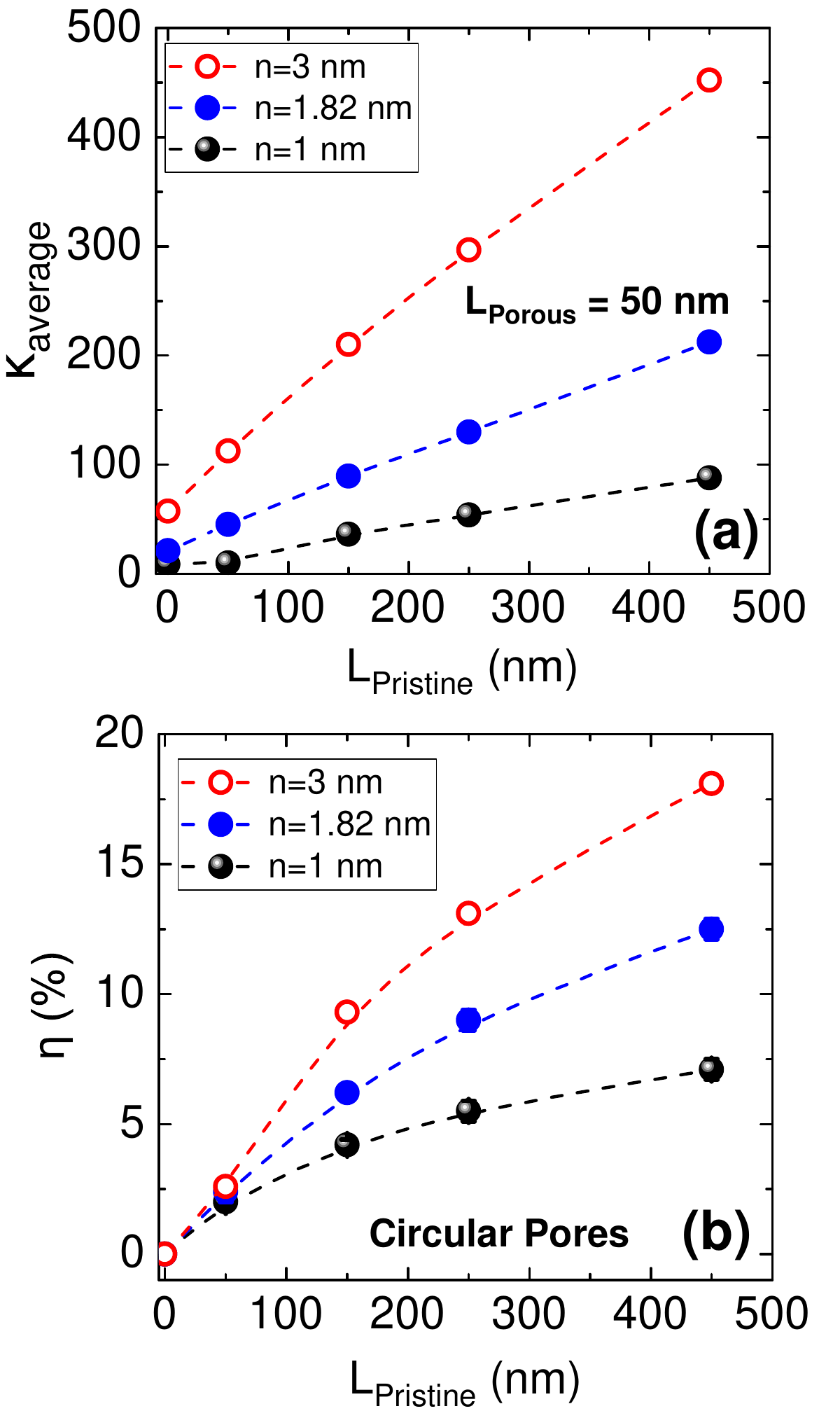}
        \caption{\label{fig:Porosity} The effect of pore neck size $n$ on the  dependence of (a) the thermal conductivity $\kappa(L)$ and (b) the thermal rectification ratio $\eta(L)$ of partially perforated graphene nanostructures on the pristine side length $L_{pristine}$. The structures studied had circular pores and different neck sizes ($n$=1, 1.82 and 3~nm) and a constant perforated length $L_{porous}$=50 nm, corresponding to the configuration of Fig.~\ref{fig:Config}e. Although the errors were also calculated, the error bars are smaller than the symbols used, and therefore are not visible. For $\kappa$ only the mean value is displayed.}
    \end{center}
\end{figure}

In order to obtain rectification from perforated graphene structures, we first turned to creating the necessary asymmetry through the introduction of pores only on one side of the structure (partially perforated graphene) and studied the effects of the pores' geometric parameters on the thermal conductivity $\kappa$ and rectification ratio $\eta$ of these systems. In this part of our study we modelled partially perforated graphene nanostructures of a constant porous length of $L_{porous}$=50~nm and varying the length of the pristine side $L_{pristine}$ between 50-450~nm. This was initially done to probe $L_{pristine}$ as a parameter, but it was later revealed by the analysis with the macroscopic R-Series model (see Section~\ref{subsect:R_Series}) that the resistance-matching condition for maximum rectification is achieved for very large $L_{pristine}$. The porous part consisted of either circular or triangular pores with constant $n_x$=$n_y$=$n$ and $n+D$= 5~nm (no spatial variation), similar to the systems shown in Figs.~\ref{fig:Config}e-\ref{fig:Config}h.

First, in an attempt to examine the influence of the pore neck/diameter size on $\kappa$ and $\eta$, we focused on circular holes having different neck sizes $n$. We studied three different values of $n$ and calculated $\kappa(L)$ and $\eta(L)$ as a function of system length $L$. We investigated $n$= 1, 1.82, 3~nm, corresponding to porous graphene with $D$= 4, 3.18, 2~nm and $p$=50\%, 32\% and 12.5\% respectively. The results of these calculations are presented in Figure~\ref{fig:Porosity}. Several notable results are to be highlighted here. First of all, in all partially perforated configurations treated in this study, the `easy' heat flow direction is exclusively from the porous to the pristine side. Second, in all the above cases, both $\kappa$ (Fig.~\ref{fig:Porosity}a) and the thermal rectification ratio $\eta$ (Fig.~\ref{fig:Porosity}b) increases with $L_{pristine}$, although contrary to either the entirely porous or entire pristine case (see Fig~SM-5), $\kappa(L)$ appears almost linear. This result will be further clarified in the next subsection, where it will be understood through the lens of the R-series model. Third, both $\kappa$ and $\eta$ increase with increasing the pore necks $n$ for all $L_{porous}$. The increase in $\kappa$ can be directly and intuitively understood as being due to the increase in the width of the phonon pathways, associated with $n$. The increase of $\eta$ with $n$, however, is not evident intuitively and will also be explained through the lens of the R-series model as being due to better resistance-matching with the pristine part of the heterostructure. The maximum attained value for $\eta$ was $\sim$18.6\%, for $n$=3~nm and $L_{pristine}$= 450~nm, which was also the maximum rectification ratio found in the present study.

\begin{figure}[tbp]
    \begin{center}
        \includegraphics[width=0.8\columnwidth]{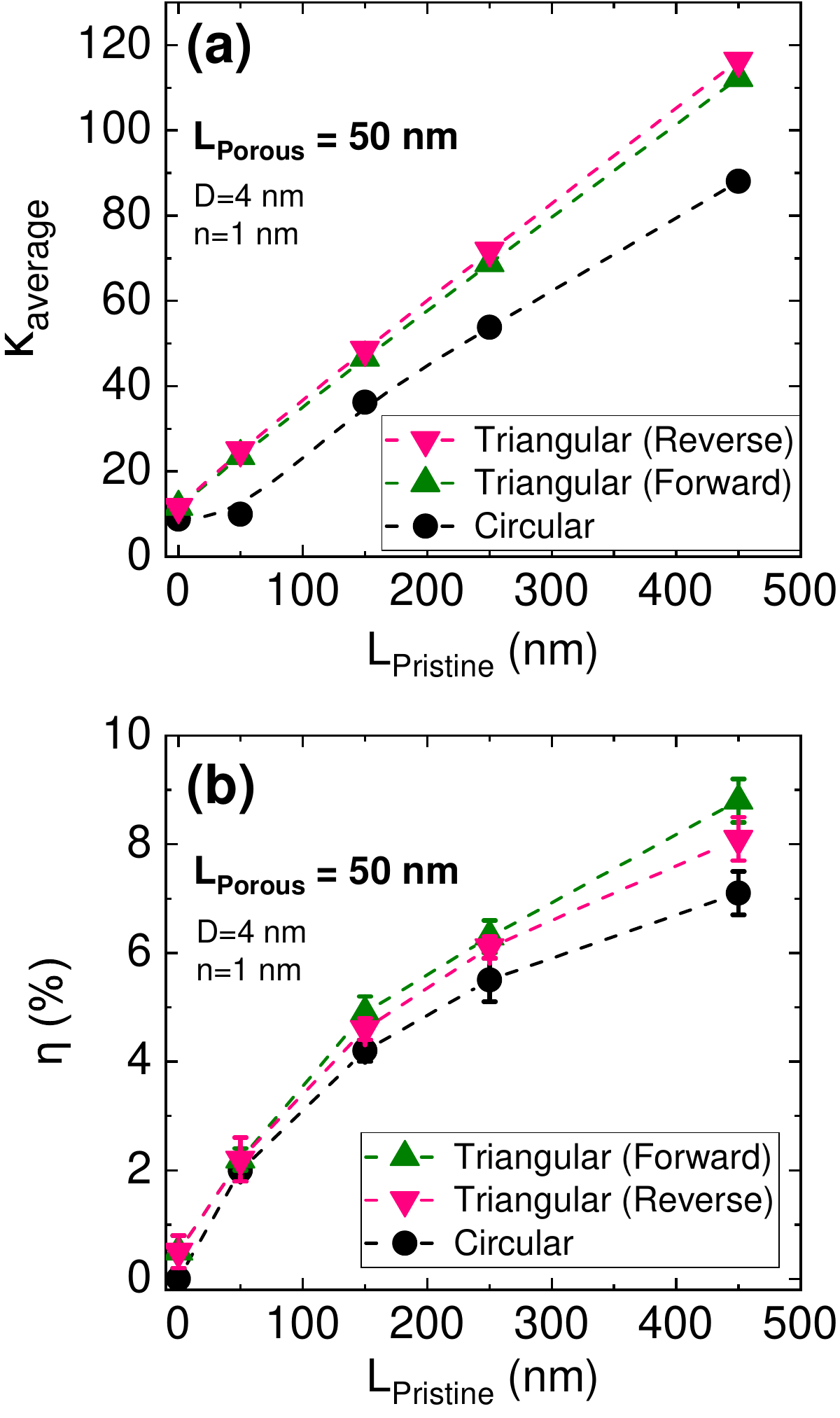}
        \caption{\label{fig:Shape} The effect of pore shape and orientation on the dependence of (a) the thermal conductivity $\kappa(L)$ and (b) the thermal rectification ratio $\eta(L)$ of partially perforated graphene nanostructures on the pristine side length $L_{pristine}$. The structures studied have circular pores, triangular pores with their apexes oriented towards (\textit{`forward'}) and away from (\textit{`reverse'}) the pristine part, all having the same pore diameters $D$=4~nm and necks $n$=1~nm and a constant porous length $L_{porous}$=50~nm (Figs~\ref{fig:Config}e-g). For $\eta$ the error bars are also displayed, while for $\kappa$ only the mean value is displayed.}
    \end{center}
\end{figure}

Next, we investigated the possible effect of the pore shape in the thermal conductivity and rectification of the nanostructures. In order to do so, we have performed additional calculations comparing the thermal properties of partially perforated graphene with circular and triangular pores, all the while having the same pore diameters $D$=4~nm and necks $n$=1~nm. These correspond to the largest circular pores. This choice was made because for larger pores, the difference between shapes is more pronounced, and thus any possible effect of the shape on the thermal properties can be more easily observed. For the triangular pores, we have also examined both the \textit{`forward'} and the \textit{`reverse'} orientations of the pores (see Figs.~\ref{fig:Config}g and \ref{fig:Config}h respectively). It shall be noted that although the triangular and circular pores have the same $D$ and $n$ values, they have quite different porosities, due the shape of the pores ($p$=50\% for circular and $p$=32\% for triangular pores). The results of these calculations are displayed in Fig.~\ref{fig:Shape} and the following observations can be made.

First, the pore orientation does not seem to play a crucial role in controlling thermal rectification, at least for the pore length scales studied here. The $\kappa(L)$ and $\eta(L)$ of the systems with the \textit{`forward'} and \textit{`reverse'} orientations of the triangular pores are practically identical. Besides, at $L_{pristine}$=0~nm, the $\kappa$ and $\eta$ of both systems are identical, implying that the asymmetric phonon scattering due to the pores' triangular shape is not enough to induce rectification.

Furthermore, we can observe that differences in $\kappa(L)$ and $\eta$ of the circular and triangular systems are small, despite them having big differences in porosity (e.g. the triangular system had $\sim$20\% larger $\kappa$ and $\eta$ for $L_{pristine}$=450~nm). This becomes even more evident if we compare a system with circular and triangular pores having the same porosity $p$=32\% ($n$=1.82~nm for the circular and $n$=1~nm for the triangular); for $L_{pristine}$= 450~nm the system with the circular pores had significantly larger $\kappa$ (+88\%) and $\eta$ (+55\%).

Finally, in order to investigate a possible effect of pore alignement, we decided to repeat the above calculations with triangular pores having identical geometry ($D$=4~nm, $n$=1~nm, $p$=32\% and \textit{`forward'} orientation), but with the triangular pores in subsequent rows misaligned, as shown in Fig.~\ref{fig:Config}h. For the best case ($L_{pristine}$=450~nm), the results of these calculations showed that $\kappa$ of misaligned pores was $\sim$35\% lower (76 vs 117 W/mK) and $\eta$ was $\sim$20-25\% lower (6.1\% vs 8.1\% for \textit{`forward'} and 8.5\% for \textit{`reverse'} aligned). This result implies that pore alignment is quite efficient in controlling thermal conductivity, by modifying the effective width $n$ of the \textit{`phonon pathways'}. In turn, this again supports the argument that for pure perforated graphene porosity does not seem to be a good metric in determining neither $\kappa$ nor $\eta$, but rather it is the width of the \textit{`phonon pathways'}, here quantified by $n$, that is more crucial for determining bith thermal conductivity and rectification. The effect of alignment on rectification, however is much less pronounced and, as will be shown in the next subsection, the effect is primarily due to the reduction in $\kappa$ and thus the resistance-matching with the pristine part, as explained by the R-series model.

\begin{figure}[!tp]
    \begin{center}
        \includegraphics[width=1\columnwidth]{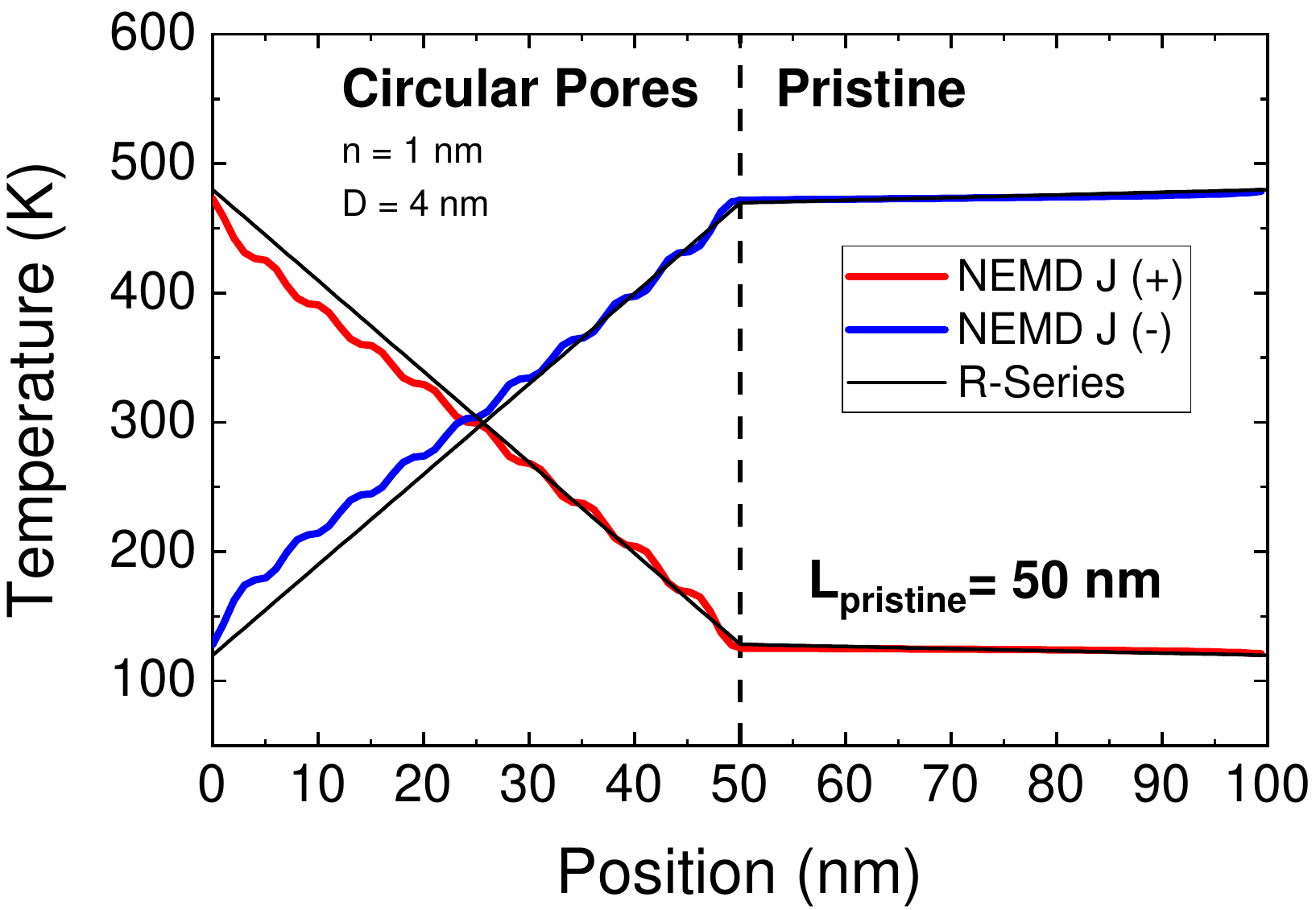}
        \caption{\label{fig:T_Profile} The temperature profile $T(x)$ from of a porous-pristine heterostructure with circular holes and $n$=4~nm and $D$=1~nm, for the forward (heat flow from porous to pristine) and the reverse  heat flux direction (heat flow from pristine to porous). The $T(x)$ presented are the averages over 5 independent NEMD simulations for each direction. The system lengths are $L_{porous}$=$L_{pristine}$=50~nm and the interface is denoted with a vertical dashed line. Straight black lines correspond to the T(x) predicted by the R-Series model with no interfacial resistance included.}
    \end{center}
\end{figure}

\subsection{The R-series Model}

In this section we shall attempt to explain the trends in $\kappa(L)$ and $\eta(L)$ of the partially perforated graphene structures observed in the previous subsection, by comparing the theoretical predictions of the macroscopic R-series model with the results obtained with NEMD. Within the scope of this model, the porous and the pristine parts of the heterostructure are treated as individual autonomous systems, which are then connected \textit{`in series'}, while neglecting any interfacial thermal resistance between them. This is actually a fair approximation, since, although there is a mesoscopic interface between perforated and pristine graphene, the main thermal resistance occurs at the pores level of the perforated side. This is also supported by an inspection of the temperature profile of a typical NEMD simulation of a heterostructure, where it is evident that the temperature profile exibits a negligible temperature jump $\Delta$T$_{i}$ across the porous-pristine interface, as seen for example in Fig.~\ref{fig:T_Profile} for a circular pore system with $n$=1~nm and $L_{porous}$=$L_{pristine}$=50~nm (see also Figs.~SM-1 and SM-2 for calculations of $\Delta$T$_{i}$ for other configurations as well).

\begin{figure}[!tbp]
    \begin{center}
        \includegraphics[width=0.8\columnwidth]{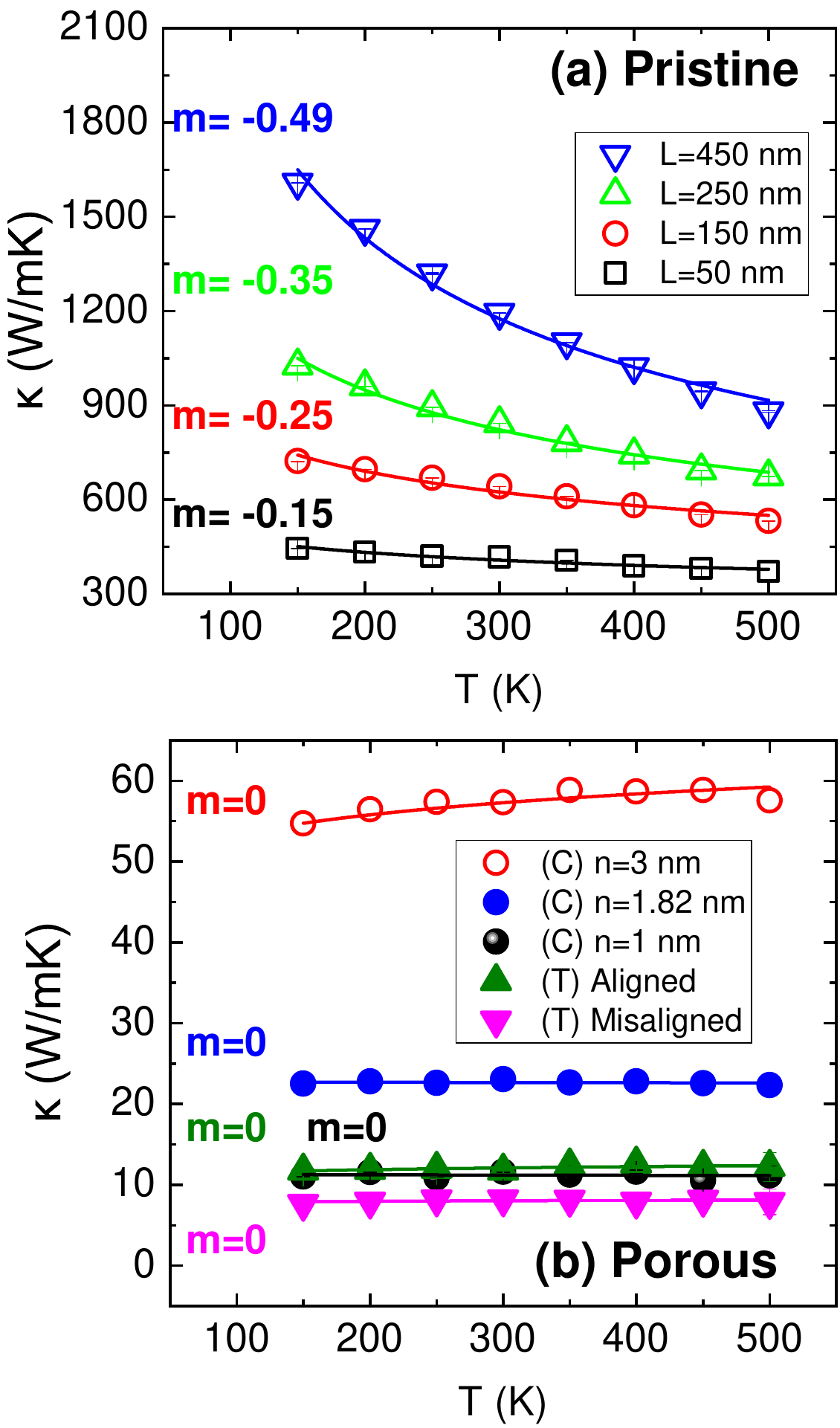}
        \caption{\label{fig:T_Dependence} The temperature dependence of the thermal conductivity $\kappa(T)$ of all the graphene-based systems treated in this work with NEMD. (a) Pristine graphene with $L$=50, 150, 250 and 450~nm, and (b) porous graphene with circular pores and $n$=1, 1.82 and 3 nm, and with aligned and misaligned triangular pores (\textit{`forward'} orientation), having $n$=1~nm. The solid lines represent the power law fits of eq.~(\ref{eq:Power_Law}) to the NEMD results.}
    \end{center}
\end{figure}

We first need to determine the temperature dependence of the thermal conductivities of the materials composing the heterostructures, and determine $\kappa_0$ and $m$ of eq.~(\ref{eq:Power_Law}) for each material. For this we have calculated $\kappa(T)$ of the pristine and porous graphene systems studied in this work, namely for pristine graphene with $L$=50, 150, 250 and 450~nm, as well as of porous graphene with circular pores and $n$=1, 1.82 and 3 nm, and with aligned and misaligned triangular pores (\textit{`forward'} orientation), having $n$=1~nm. The NEMD values of $\kappa(T)$ for each individual graphene system were also checked for possible variations with $\Delta$T (ranging from 20-200~K) due to the existence of the pores, which could potentially lead to interfacial effects within the porous material, but no such variations were observed (see Fig.~SM-4). The results of these calculations are presented in Fig.~\ref{fig:T_Dependence}a (pristine systems) and Fig.~\ref{fig:T_Dependence}b (porous systems).

From these results we can immediately observe that $\kappa(T)$ of pristine graphene decreasing with $T$ due to $ph$-$ph$ scattering, while the power law exponent increases in absolute value with the system length $L$. For a sufficiently large graphene system ($L \gg l_{MFP}$), where boundary scattering is  insignificant and the dominant phonon scattering mechanisms is Ummklap $ph$-$ph$ scattering, we expect a power law exponent, (depending on the relative strength of 3- and 4-$ph$ processes on the phonon relaxation times~\cite{Balkanski1983, Klemens1966}). Ab initio BTE calculations including 3$ph$+4$ph$ contributions have suggested an exponent of $m$= -1.3 for the intrinsic $\kappa$ of graphene (no boundary scattering, infinite length)~\cite{Han2023}. For the lengths studied ($L\leq$450~nm), $|m|<$0.5, as we are in the classical regime and at finite lengths. Therefore, by further increasing $L$, the exponent should increase along (in absolute value). From MD simulations performed on pristine graphene, it was suggested that this diffusive limit of $\kappa(L)$ is reached for $L\sim$100~$\mu$m~\cite{Barbarino2015,Park2013}. 

Concerning perforated graphene, $\kappa$ is practically independent of $T$ (see Fig.~\ref{fig:T_Dependence}b). This is more or less expected from MD simulations~\cite{Donadio2010}, since MD does not capture the correct quantum statistics, and thus the dominant mechanism that determines $\kappa$ is the boundary scattering from the pores, which is independent of $T$. Although, $\kappa$ can be made to increase with $T$ by nanostructuraton, even at higher temperatures~\cite{Shiomi2018}, an appropriate temperature rescaling scheme that incorporates Bose-Einstein statistics can be applied to MD simulations in order to correctly predict $\kappa(T)$ at this regime,~\cite{Lee1991}. In any case, the NEMD simulations probably underestimate the power law exponents for both pristine and porous, and these effects can partially cancel out (see eq.~(\ref{eq:Linear_R-Series})).

\begin{figure}[!tbp]
    \begin{center}
        \includegraphics[width=0.8\columnwidth]{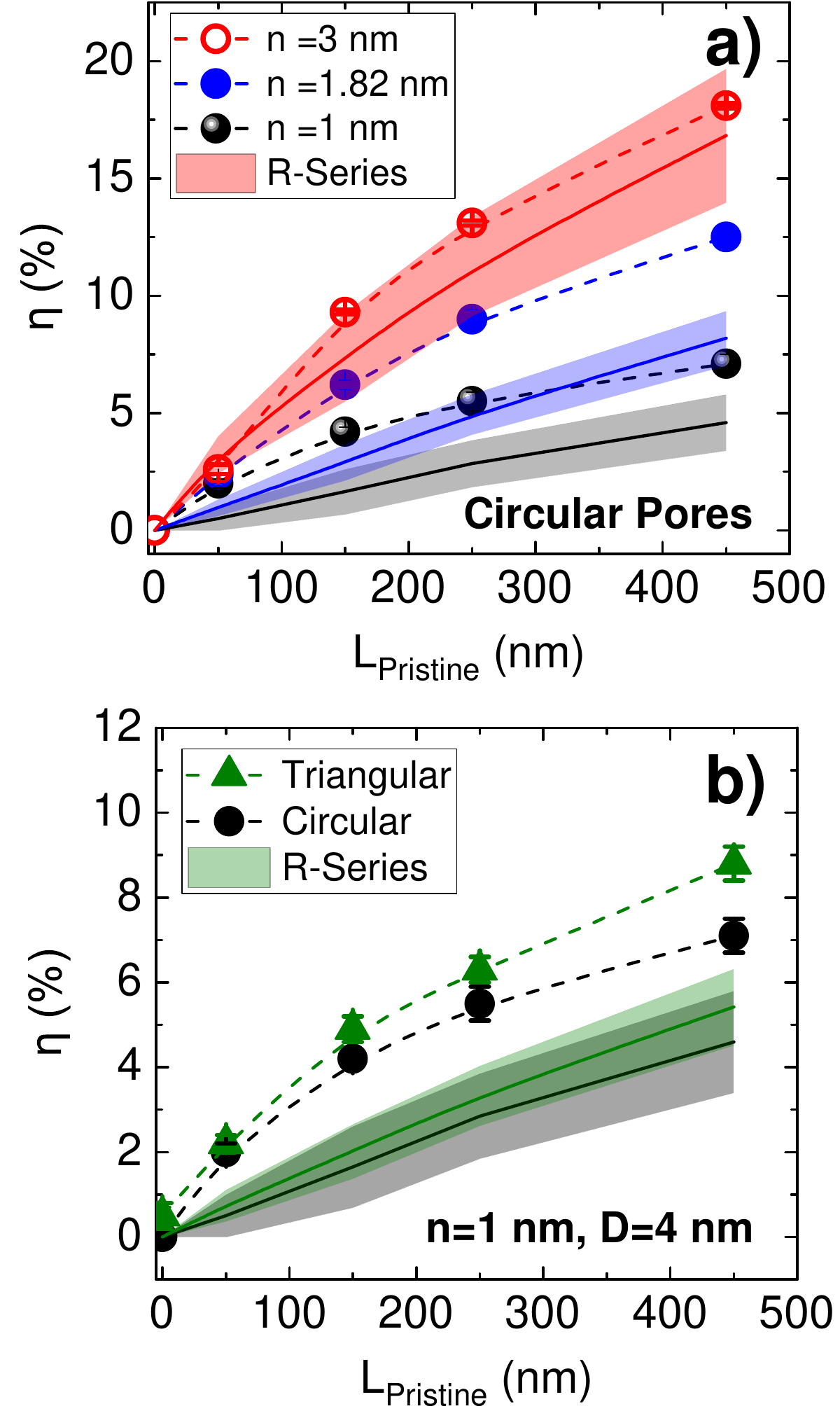}
        \caption{\label{fig:R_Series} The R-series model (solid lines) applied to the systems studied in this work, in comparison to the NEMD results (symbols) for the thermal rectification for (a) circular pores with $n$=1, 1.82 and 3~nm and (b) circular and triangular pores with $n$=1~nm. The systems studied have a constant porous length $L_{porous}$=50~nm and varying pristine side lengths $L_{pristine}$=50-450~nm and correspond to Figs~\ref{fig:Config}e-f. For the R-series results, the shaded areas represent the uncertainties in the fitting parameters $\left(\kappa_0, m\right)$, while for the NEMD results the error bars are also displayed. Some error bars for the NEMD results are smaller than the symbols used, and therefore, not visible.} 
    \end{center}
\end{figure}

We have used the obtained power-law parameters $(\kappa_0, m)$ for each system, and the imposed temperature bias $\Delta T$=360~K used in the NEMD simulations, to calculate the reduced heat current $q$ for each direction of the heat flow, by solving the system of eqs.~(\ref{eq:R_Series_Q}), as discussed in Section~\ref{subsect:R_Series}. Special care was taken to include the propagation of the uncertainties of the fitting of eq.~(\ref{eq:Power_Law}) to the final reults for $\eta$\%. which can be sometimes significant. The results of these calculations, along with the uncertainties, are presented in Fig.~\ref{fig:R_Series}, where we compare them with the NEMD results for the thermal rectification of the systems studied. This macroscopic model appears to describe correctly the qualitative trends of the NEMD rectification results, but fails quantitatively, especially for larger pores.

In accordance with the NEMD results, the easy axis is from the porous to the pristine side, rectification increases with increasing $L_{pristine}$, and that decreasing the pore neck $n$ increases $\eta$, up to the length scales studied in this work. We can further observe some significant uncertainties in the predictions of the model, as witnessed by the shaded areas in Fig.~\ref{fig:R_Series}. These significant uncertainties are primarily due to the fitting errors of the power law exponent $m$, whose mean values are close to zero for all porous systems. We can nonetheless observe that despite this large uncertainty, the R-Series model qualitative predictions are still valid.

However, for larger pores, the model significantly underestimates TR even with the uncertainty taken into account, irrespective of shape (e.g. Fig.~\ref{fig:R_Series}b). We claim that the origin of this quantitative discrepancy lies with the fact that the R-Series model is macroscopic and assumes that the thermal properties of each material are local and continuously changing within the temperature profile $T(x)$. However, the $\kappa(T)$ of the porous graphene systems calculated by NEMD are strictly global properties and therefore, we cannot attribute a local value to the thermal conductivity that is equal to $\kappa_{local}=\kappa(T(x))$, especially for the systems with larger pores which are highly inhomogenous. This constraint can be relaxed for the smaller pores, where the material becomes more homogeneous at the length scales of the total system lengths, and the assumptions of the R-series model become more accurate.

\begin{figure}[!tbp]
    \begin{center}
        \includegraphics[width=0.99\columnwidth]{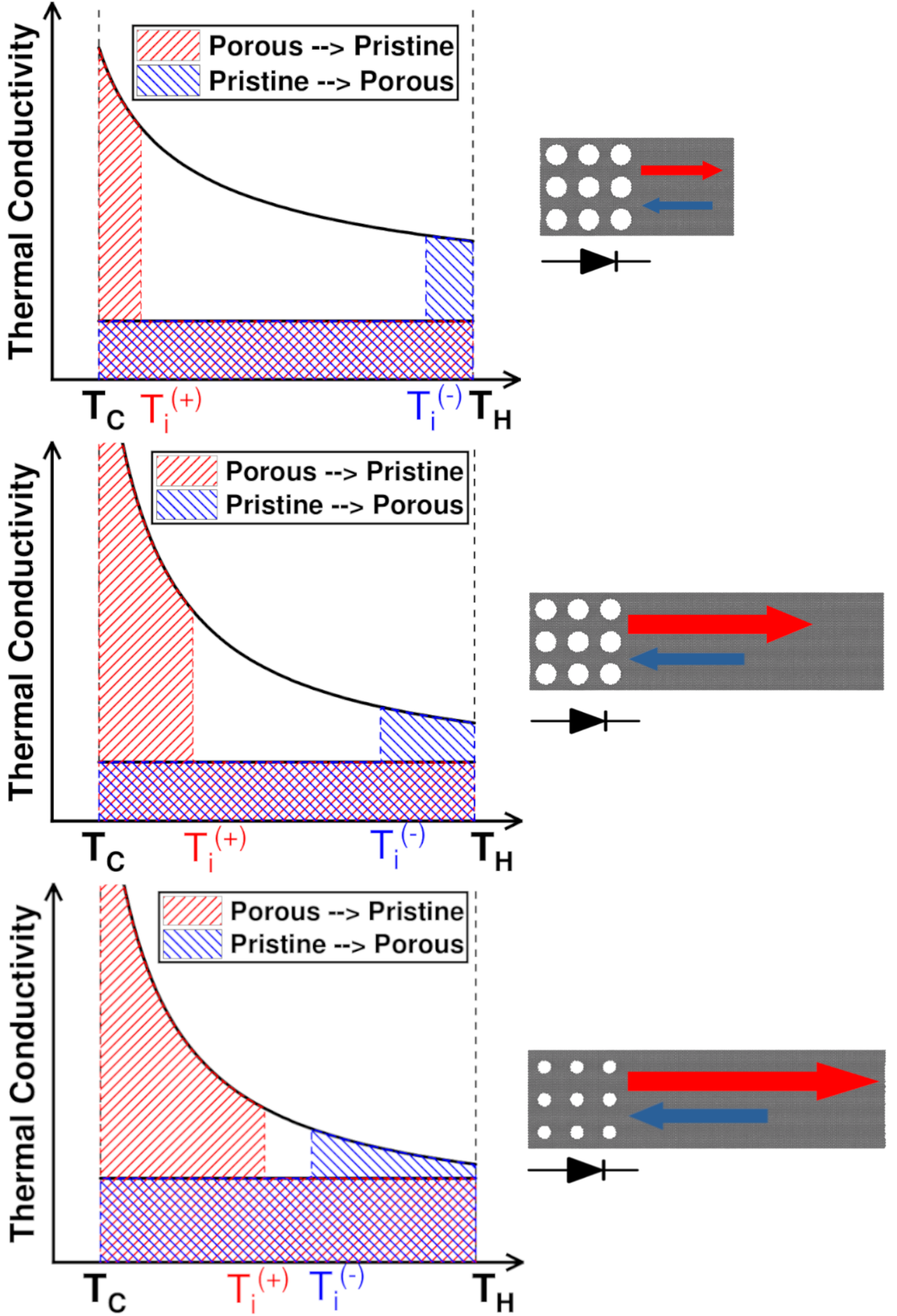}
        \caption{\label{fig:Vis_Abst} A visual schematic of the thermal rectification trends in partially perforated graphene, explained by the R-Series model. The heat flux is equal to the area of $\kappa(T)$ between $T_C$ or $T_H$ and the interface temperature $T_i$, which is potentially different for each direction. The resistance matching criterion ($\rho \rightarrow$1) results in $T_i^+ \rightarrow T_i^-$, while the exponent mismatching criterion ($\Delta m \gg 0$) makes the two fluxes (red vs blue areas) very dissimilar. (\textit{Top}) The direction of heat flow from the porous to the pristine (J$^+$: red) is always larger than from the pristine to the porous (J$^-$: blue). (\textit{Middle}) Increasing $L_{pristine}$ increases $R_{pristine}$ and decreases $m_{pristine}$, thus satisfying both criteria. (\textit{Bottom}) Decreasing the pore size decreases $R_{porous}$, thus satisfying the first criterion.}
    \end{center}
\end{figure}

Since the model is qualitatively accurate, we can rely on it to further enhance the rectification performance. For this, we can make use of the linearized version of eqs.~\ref{eq:R_Series_Q}, which provides a lower limit to the full solution and is valid for small $\Delta$T~\cite{Dames2009}:
\begin{eqnarray}
    \eta_{linear}=\frac{m_1-m_2}{\left(\rho^{1/2}+\rho^{-1/2}\right)^2}\left(\frac{\Delta T}{T_0}\right) \label{eq:Linear_R-Series}
\end{eqnarray}
Although the $\Delta T$ used is not small in our case, the linearized version provides a similar qualitative trend to the full solution~\cite{Dames2009}, as well as simple strategies towards maximization of $\eta$. One can thus directly see that rectification increases with the relative temperature bias $\Delta$T/T$_0$, which is a concrete manifestation of the necessary condition of non-linearity for thermal rectification, first mentioned in the Introduction. Next, for a given $\Delta$T, in order to maximize $\eta$, one sees that $\rho$ should be as close to 1 as possible (resistance matching criterion) and the exponent difference $\Delta m$ should be as high as possible (exponent mismatching criterion). These are the two most important criteria for maximizing $\eta$ in the R-series model~\cite{Dames2009}. The application of these two criteria to explain the rectification trends in the systems studied in this work is presented visually in Fig.~\ref{fig:Vis_Abst}. According to the model, the heat flow is equal to the area of $\int_{T_C}^{T_i}\kappa_1(T)dT$= $\int_{T_i}^{T_H}\kappa_2(T)dT$ (shaded areas in Fig.~\ref{fig:Vis_Abst}) between $T_C$ or $T_H$ and the interface temperature $T_i$, which is potentially different for each direction. The resistance matching criterion ($\rho \rightarrow$ 1) implies that $T_i^+ \rightarrow T_i^-$, while the exponent mismatching criterion ($\Delta m \gg 0$) makes the two fluxes (red vs blue areas in Fig.~\ref{fig:Vis_Abst}) very dissimilar. 

For the first criterion (resistance matching), we need the porous graphene to have a $\kappa$ as high as possible, in order to match the (low) resistance of the pristine. This can be done by decreasing the size of the pores and it in fact explains why the circular $n$=1~nm system has the highest $\eta$ (Fig.~\ref{fig:Vis_Abst} \textit{Top vs Bottom}). Alternatively, increasing $L_{pristine}$ increases $R_{pristine}$, thus matching the (high) resistance of the porous side (Fig.~\ref{fig:Vis_Abst} \textit{Top vs Middle}). This increase of $R_{pristine}$ can be expected since $R(L)=\frac{L}{\kappa(L)}$ and the thermal conductivity of pristine graphene follows a logarithmic law $\kappa \sim log(L)$~\cite{Barbarino2015,Xu_LogL2014,Park2013} below the diffusive limit of $L\sim$100~$\mu$m, thus $\kappa_{pristine}$ always increases slower than $L$. However, after a certain threshold for $L_{pristine}$, the matching condition will be reversed, $R_{pristine} \geq R_{porous}$ and thus after this threshold, $\eta$ will start reducing again. On the other hand, regarding the second criterion (exponent mismatching), increasing $L_{pristine}$ leads to decreasing the pristine side exponent $m_2$ towards more negative values, enhancing $\Delta m$ and thus $\eta$. Furthermore, related to this, the pore size cannot be decreased indefinitely to maximize resistance matching, since after a certain threshold the boundary scattering due to the pores will stop being dominant and the porous side exponent $m_1$ will become negative, thus reducing $\eta$. The above rectification trends are visually summarized in Fig.~\ref{fig:Vis_Abst}.

Based upon the previous discussion, we can use the linearized R-series model to make predictions. As an example, an increase in $L_{porous}$ will produce the opposite effect in MD calculations, as both the porous side exponent $m_1$ and $\kappa_{porous}(L)$ are expected to be constant independently of $L$, thus increasing $R_{porous}$ with $L$ and drifting away from the $\rho$=1 condition. Likewise, we may predict that according to eq.~\ref{eq:Linear_R-Series}, by increasing the mean temperature $T_0$ of the system (while keeping the ratio $\Delta T / T_0$ the same), $\eta$ will further increase, since for perforated graphene $\kappa_0$ is practically constant with temperature, while for pristine graphene $\kappa_0$ decreases with $T$, thus having a higher mean resistance $R_0$ and thus better matching the resistance of the porous side. 

\subsection{Insufficiency of inhomogeneous perforation to produce thermal rectification without pristine}
We close our analysis by investigating the effects of inhomogenous perforation on the thermal conductivity and rectification of monolayer graphene, this time without combination with pristine. The motivation for this last part was primarily to determine whether a non-uniform perforation pattern would induce sufficient asymmetry in order to observe significant thermal rectification. The configurations studied are also depicted in Fig.~\ref{fig:Config}b-d (for more details about the structures see Supplementary Material Section 4). However, in all the configurations studied, no thermal rectification was observed above the uncertainty (1\%). The results nonetheless add insight to the discussion, as it is found that although no rectification is found, the structure where the $n_y$ is not constant along the heat flow (Fig.~\ref{fig:Config}c) gave systematically lower thermal conductivity than the other two, which had constant $n_y$. This is in line with the discussion in Section~\ref{subsect:Heteros}, that the width of the `\textit{phonon pathway}' is the most crucial factor in determining the thermal conductivity of porous graphene systems. Furthermore, a comparison with pristine graphene has shown that perforation reduces graphene's thermal conductivity by up to 85\% (see Fig. SM-5). This is a significant reduction, however it is still much less than the drop of $\kappa$ due to single-vacancy defects, which can reach up to 99\%~\cite{Zhao2015,Haskins2011}. This means that pores are more efficient than vacancies to produce thermal rectification for the same porosity, as they do not significantly reduce $\kappa$, thus matching the criterion $\rho$=1 while still maintaining $m$=0 (R-Series model criteria).

\section{Conclusions\label{sec:Conclusions}}

In this work we have performed Non-Equilibrium MD simulations of heat transport in partially perforated monolayer graphene nanostructures, in order to calculate the thermal conductivity $\kappa$ and the thermal rectification ratio $\eta$ of these systems. In all cases rectification was observed, with a maximum value of $\eta$= 18.5\% and with heat preferentially flowing from the porous to the pristine sides. Thermal rectification $\eta$ and conductivity $\kappa$ increase with the length of the pristine region ($L_{\text{pristine}}$) and the pore neck size $n_y$ (perpendicular to the heat flow). While pore shape and alignment minimally influence rectification, misaligned pores disrupt phonon pathways, further reducing $\kappa$. The maximum rectification was observed in systems with the smallest pores ($D$=2~nm) and longest pristine regions ($L_{pristine}$=450~nm). 

Perforation of pristine graphene alters its phonon spectrum by modifying pre-existing bulk modes and introducing localized defect modes, particularly around or between pores. These defect modes exhibit low participation ratios $p_k$, with out-of-plane vibrations showing even lower $p_k$ than in-plane modes. A prominent defect mode at 520~cm$^{-1}$, corresponding to radial vibrations of under-coordinated pore-edge atoms, produces a distinct peak in the density of states (DOS). This peak's intensity scales with porosity due to the increasing number of edge atoms. Notably, the DOS depends solely on porosity, not pore shape or alignment. Perforation also reduces thermal conductivity $\kappa$ significantly (up to 85\% for 13\% porosity), though less drastically than randomly distributed defects (e.g., single vacancies). Crucially, $\kappa$ is governed not by porosity alone but principally by the width of the `\textit{phonon pathways}', defined as the inter-pore neck dimension $n_y$, perpendicular to the heat flow.

In order to shed light into the mechanism of thermal rectification in the systems studied above, we employed the macroscopic R-Series model, which treats the pristine and perforated parts of the nanostructure as individual thermal resistances connected in series, attributes rectification in such a structure to the difference in the temperature dependence of $\kappa$ of the two constituent materials, while neglecting any interfacial effects. The model used further assumes that $\kappa(T)$ is given by a power law $\kappa(T) \sim \kappa_0 (T/T_0)^m$, which was also calculated using MD. Our results reveal that $\kappa(T)$ in pristine graphene decreases with temperature due to Umklapp phonon-phonon scattering, whereas porous graphene's $\kappa(T)$ remains temperature-independent due to the dominance of boundary scattering and the absence of quantum effects.

The R-series model captures correctly the qualitative rectification trends, namely that the rectification ratio $\eta$ increases with increasing $L_{pristine}$ and decreasing $D$. This is a strong confirmation that the dominant mechanism of thermal rectification in partially perforated graphene nanostructures is primarily due to the difference in the temperature dependence $\kappa(T)$ of pristine and perforated graphene. The model also provides a lower limit to the true rectification ratio, and can be used to determine maximization strategies. In all cases, according to the model, the maximum value of $\eta$ is obtained by having the two materials comprising the heterostructure have matching thermal resistances and mismatching exponents of the temperature dependence of $\kappa$. Therefore, to enhance $\eta$ in our case, possible strategies may include minimizing pore size, optimizing $L_{\text{pristine}}$, reducing $L_{\text{porous}}$, and prioritizing periodically arranged pores over randomly distributed defects, though limits depend on system specifics. Ultimately, maximizing rectification hinges on balancing resistance matching and mismatched power-law exponents between pristine and perforated regions. Nonetheless, care must be taken in regards to the quantitative predictions, since the model underestimates $\eta$ for largely inhomogenous systems like the ones with larger pores, as it assumes homogeneous thermal properties, an assumption which becomes increasingly invalid for larger pores. For accurate quantitative results, NEMD calculations are necessary. \\

\section*{\label{sec:Vide} Acknowledgments}
Authors acknowledge the support of the French National Research Agency (ANR) through the grant ANR-24-CE50-2146 (project `BF2D'). The major part of the calculations presented in this work was performed in the \textit{Jean-Zay} CPU array of the IDRIS High Performance Computing facilities (HPC) under the allocation 2022-AD010913913 made by
GENCI. The authors would also like to thank Dr. Roman Anufriev and Dr. Olivier Bourgeois for the fruitful discussions on the R-series model and its application on the partially perforated graphene nanostructures, as well as Dr. Nedjma Bendiab for the helpful discussions on the experimental aspects of graphene perforation.

\end{document}